# Models of the Night-Sky Brightness and the Efficiency of Searching for Exoplanets with the Microlensing Method

S. I. Ipatov*

*Vernadsky Institute of Geochemistry and Analytical Chemistry, Russian Academy of Sciences, Moscow, 119991 Russia*
*e-mail: siipatov@hotmail.com*

**Abstract**—We analyze photometric observations of stars, which experienced microlensing events at the considered time, in order to compare the efficiency of detecting exoplanets in observations performed at thirteen different telescopes and with several approaches to the selection of observable events. In constructing an algorithm of the optimal selection of targets for these observations and in comparing the detection efficiencies for several telescopes, we considered models of the night-sky brightness that satisfy the data of infrared observations carried out in 2011 with the Optical Gravitational Lensing Experiment (OGLE) telescope and the RoboNet telescopes (FTS, FTN, and LT) used to search for planets with the microlensing method. The considered models of the night-sky brightness can be used for various observations (not only microlensing events). The time intervals, during which microlensing events can be observed, were determined with accounting for the positions of the Sun and the Moon and the other constraints on the telescope pointing. Our algorithm allows us to determine the already known microlensing events that are accessible for observation with a particular telescope and to select targets, for which the probability of detecting an exoplanet is maximal. The events, which would maximize the probability of detecting exoplanets, were selected for observations. The probability of detecting an exoplanet is usually proportional to the mirror diameter of a telescope. Telescopes with a wider field of view, such as the OGLE, are more effective in finding new microlensing events. To observe different microlensing events, it is usually better to use different nearby telescopes. However, all such telescopes are often better to use for observing the same event in those relatively short time intervals that correspond to the peak brightness of the event.

## INTRODUCTION

At present, more than 5700 planets have been found near other stars (Ananyeva et al., 2022; Marov and Shevchenko, 2020, 2022). As of April 2025, 1112 exoplanets have been discovered by the Doppler method (from the radial velocities of stars), 4360 exoplanets by the method of transits (from changes in the brightness of stars), 235 exoplanets by microlensing, 83 exoplanets by direct observations, and 5 exoplanets by astrometric methods (from measurements of the star's position). Current data on the number of exoplanets discovered by various methods are available at https://exoplanets.nasa.gov/alien-worlds/ways-to-find-a-planet/.

The gravitational microlensing method is based on ability of gravitation to change the intensity of light from a star and to bend light beams. This effect was first described by A. Einstein. When a beam of light from a source star passes a lensing star on its way to the Earth, its trajectory is curved, which can enhance the observed brightness of the source star. This flash in the intensity of light is called a microlensing event. An additional smaller flash in the intensity of light can occur, if the lensing star has a planet. This method is used in searching for not only exoplanets, but also other celestial objects. For example, Zakharov and Sazhin (1998) outline the basics of the standard theory of the microlensing of stars.

The microlensing effect is characterized by the Einstein angular radius $\theta_E = (4GM(D_L^{-1} - D_S^{-1})/c^2)^{1/2}$, where $M$ is the mass of a lensing star, $D_L$ is the distance from the observer to the lensing star, $D_S$ is the distance from the observer to the light source, $G$ is the gravitational constant, and $c$ is the velocity of light. Although a microlensing event has no clear beginning or end, it is generally considered to last as long as the angular distance between the source and the lens is less than $\theta_E$. The duration of the event is determined by the time it takes for the apparent motion of the lens across the sky to cover the angular distance $\theta_E$. During the microlensing phenomenon, the brightness of the source is amplified by a factor of $A$: ($A(u) = (u^2 + 2)u^{-1}(u^2 + 4)^{-1/2}$), where $u$ is the angular distance between the lens and the source divided by $\theta_E$. If the values of $u$ are small, $A(u)$ is proportional to $u^{-1}$, i.e., proportional to $\theta_E$, and, hence, proportional to the square root of the mass of a lensing star. Real objects are not point sources, and the effects associated with a finite size of the source constrain the $A(u)$ values of the source brightness at small $u$. However, some microlensing events may cause the enhancement in brightness by a factor of hundreds.

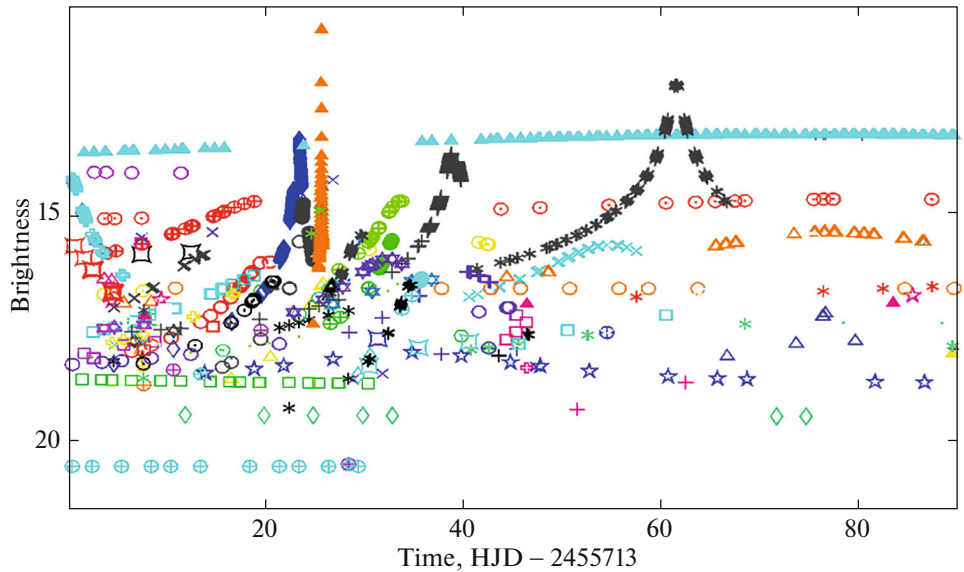

**Fig. 1.** Light curves (variations in brightness of stars) for microlensing events selected to be observed with the FTS at real moments of peaks in the light curves. The analyzed microlensing events include those of 500 events (110 501–111 000), which occurred for the 90 days considered. The values of brightness and time are expressed in stellar magnitudes and days, respectively.

One of the advantages of the microlensing method is that it allows objects emitting little light to be studied. Consequently, this method can be used to search for small planets and to detect planets that are farther away from the parent star than the planets detectable with the most other methods. Specifically, it can be used to find Earth-like planets that may support life. About the search for planets by microlensing and about the method itself, the interested reader may refer to, for example, Dominik (2010, 2012), Dominik et al. (2010), Gould (2000), Horne et al. (2009), Hundertmark et al. (2018), Street et al. (2013), and Tsapras et al. (2014). During a microlensing event, the brightness of the star increases and then returns to the previous level. Examples of changes in the brightness of stars during microlensing events are shown in Figs. 1 and 2. The time in Figs. 1, 2, and 9 is given at an interval of 90 days from the Heliocentric Julian Day (HJD) indicated in these diagrams. As can be seen from Figs. 1 and 2, the interval of changes in the brightness of stars can greatly vary from one microlensing event to another; moreover, the fraction of events, in which the brightness of a star changes by more than three stellar magnitudes, is small. The duration of a microlensing event typically ranges from a few days to several months. Once a microlensing event has been detected, other telescopes are invoked to observe it. The purpose of the observations is to detect a small additional flash in the brightness of the star, from which one can attempt to determine the parameters of the planet near the lensing star. In the example given by Dominik (2012), the duration of a signal induced by the planet was about one day and about 1.5 hours for the planets like Jupiter and like the Earth, respectively. In order not to miss a possible flash in the brightness of a star due to the presence of a planet, one should make observations quite often. During a year of observations with all telescopes used for such a search, about 2000 microlensing events are observed in different zones of the sky. According to Hundertmark et al. (2018), we may expect that, during these roughly 2000 events, about 14–46 planets can be detected near individual stars.

In the present paper, we compare the efficiency (probability) of detecting exoplanets in photometric observations performed with different telescopes for stars that experienced microlensing events at the time moments considered. In constructing an algorithm for comparing the detection efficiencies, we considered models of the night-sky brightness that satisfy the data of observations carried out in 2011 with the Optical Gravitational Lensing Experiment (OGLE) telescope and the RoboNet telescopes (FTS, FTN, and LT). The results of these studies were briefly discussed by Ipatov and Horne (2014a, 2014b) and Ipatov et al. (2013, 2014).

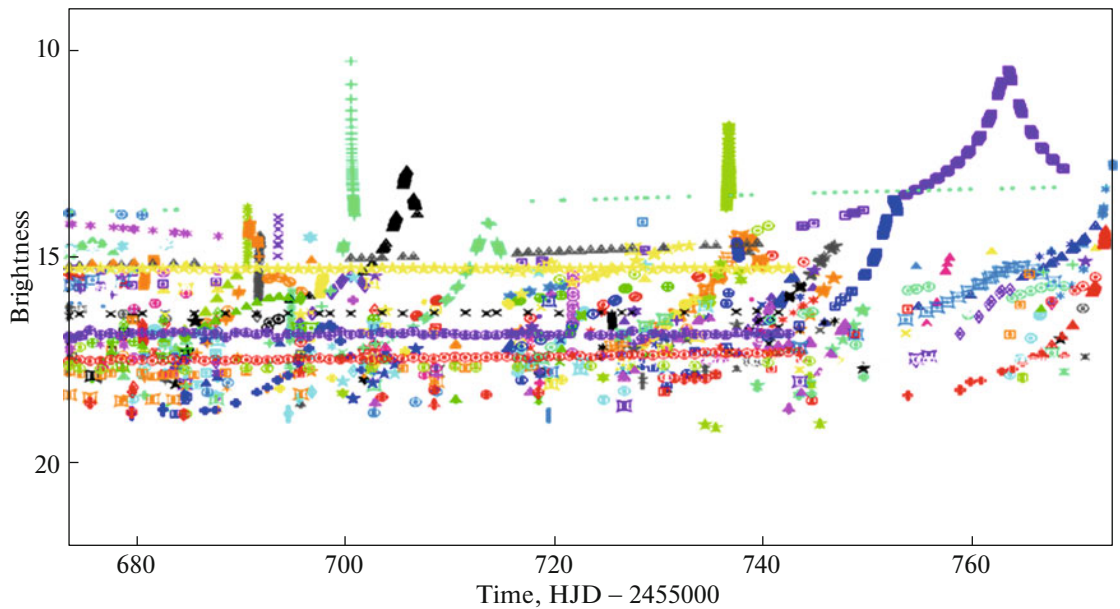

**Fig. 2.** Light curves (with error bars) for microlensing events selected to be observed with the OGLE telescope at real moments of peaks in the light curves. The analyzed microlensing events include those of 500 events (110 501–111 000), which occurred for the 90 days considered. The values of brightness and time are expressed in stellar magnitudes and days, respectively.

## AN ALGORITHM TO DETERMINE THE EXOPLANET DETECTION EFFICIENCY

With the algorithm we developed, Ipatov et al. (2013, 2014) compared the relative efficiency of detecting exoplanets in observations of known microlensing events at 13 telescopes. The comparison was performed for several models of selecting the microlensing events observed. The time intervals, in which microlensing events can be observed, are determined with accounting for the positions of the Sun and the Moon and the other constraints on the telescope pointing. Based on the approach described by Horne et al. (2009), at each time step, we calculated the detection zone and the comparative efficiency of detecting an exoplanet. At each time step, the event with the maximal detection probability was selected. We considered the following 13 telescopes, numbered from $N_t = 1$ to $N_t = 13$:

(1) 2-meter Faulkes Telescope South (FTS);
(2) 2-meter Faulkes Telescope North (FTN);
(3) 2-meter Liverpool Telescope (LT);
(4) 1.3-meter OGLE telescope;
(5–7) three 1-meter CTIO telescopes;
(8) 1-meter MDO telescope;
(9–11) three 1-meter SAAO telescopes;
(12–13) two 1-meter SSO telescopes.

Telescopes 1, 12, and 13 are at the same site—the Siding Spring Observatory (SSO) in Australia. Telescopes 5–7 are at the Cerro Tololo Inter-American Observatory (CTIO) in Chile, while three telescopes 9–11 are at the South African Astronomical Observatory (SAAO). Telescopes 2, 3, 4, and 8 are, respectively, at the observatories in Haleakela (Hawaii, USA), La Palma (the Canaries, Spain), and Las Campanas (Chile) and at the McDonald Observatory (MDO, Texas, USA).

Based on the paper by Horne et al. (2009), the detection zone was determined as an area in the plane of the lens $(x, y)$, where an anomaly in the light curve $\delta(t_i, x, y, q)$ is large enough to be detected in observations ($q$ is the ratio of the mass of the planet to the mass of the star). According to Horne et al. (2009), for the data points with a fractional accuracy $\sigma_i$ and a detection threshold $\Delta\chi^2$ at a time moment $t_i$, the detection zone $w$ is defined as $\Sigma(\delta(t_i, x, y, q)/\sigma_i)^2 > \Delta\chi^2$. The $\Delta\chi^2$ values in an interval of 25–200 correspond to a deviation of $5\sigma$ to $10\sigma$ in the light curve, if the anomaly is restricted to one data point. Horne et al. (2009, Fig. 3) show an example of the detection zone for the maximal enhancement, by a factor of five, in the brightness of the star and for $q = 0.001$. In this case, the measurement accuracy for the light curve $A(t)$ was $\sigma = 0.01(5/A^{1/2})$.

When considering a microlensing event, the planet detection probability is proportional to the zone of detecting an individual planet $w$. The signal to noise ratio is $S/N = (\Delta t/\tau)^{1/2}$, and $w = g_i \Delta t^{1/2}$, where $\Delta t$ is the exposure time and $\tau$ is the time of exposure required to achieve $S/N = 1$. The coefficient of goodness $g_i$ of the available target depends on its brightness and amplification due to microlensing, conditions of the observations (the seeing $s$, air mass $a$, and sky brightness), and characteristics of the telescope and detector.

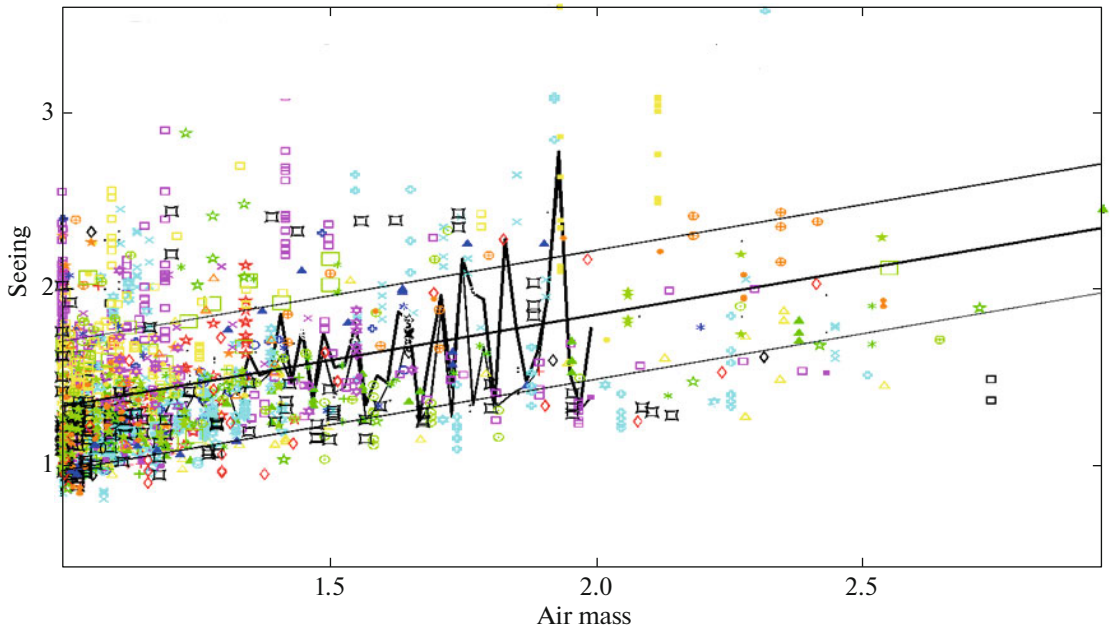

**Fig. 3.** The dependence of the seeing $s$ (the FWHM expressed in arcseconds) on the air mass $a$ (which is $a \approx \sec(z)$, where $z$ is the zenith distance) (the $\chi^2$ optimization) obtained from observations of 39 microlensing events at the FTS. The thick straight line is based on the $\chi^2$ optimization ($s = 1.334 + 0.519(a - 1)$). The thinner lines differ from the thick one by $\pm \sigma$ ($\sigma = 0.367$). The jogged lines show the mean and median values (the thicker line corresponds to the mean value).

Our program, which is based on the results of Horne et al. (2009), computes the goodness of available targets at the current time point and identifies the target that offers the largest increase in the detection zone $w$ during the exposure time. A transition to a new target occurs when the increase in the detection zone $w$ for the new target is better than that for the current target, given the telescope rotation time required to move to the new target.

Since the CCD camera requires a finite time $t_{read}$ for reading and the telescope requires a finite time $t_{slew}$ to slew from one target to another and to fix for observation of the next target, the exposure time for the target at the time of observations $t$ is $\Delta t = t - t_{slew} - n t_{read}$ (where $t_{slew} \approx 30-100$ s, $t_{read} \approx 10-20$ s, and $n$ is the number of readings). For the current target and the time step $\Delta t$, the detection zone of the $i$th event increases by $g_i[(\Delta t + t_{done})^{1/2} - t_{done}^{1/2}]$, where $t_{done}$ is the exposure time elapsed. For the next target, this zone is $g_i(\Delta t - t_{slew})^{1/2}$. To select the best event, we compared $g_i[(t_{plan} + t_{done})^{1/2} - t_{done}^{1/2}]$ for the current target to $g_i \times t_{slew}^{1/2}$ for the new target, where $t_{plan} = 2t_{slew}$. All but one of the targets require the time to change the telescope orientation before beginning the exposure.

The observability of a target is limited by its own position in the sky, as well as by the positions of the Sun and the Moon. In addition, telescopes have the pointing limitations. For example, to perform observations with the LT (http://telescope.livjm.ac.uk/), we required, among other things, that the air mass of the target not exceed 3 and the altitude of the target be between 25° and 87°, while $\cos(\theta_{Sun}) < \sin(-8.8°)$, where the solar zenith angle $\theta_{Sun}$ is the angle between the direction to the Sun and the vertical. For another telescope, there may be also limitations for the hour angle $h_\alpha$. For observations with the 1-m telescopes considered, $h_\alpha$ should be between $-5$ and $5$ h.

In the algorithm, the following input parameters of the events are used: the direct ascension and declination, the time, for which the lens is moved by the length of the Einstein ring radius, the intrinsic light flux from the source star observed, the background flux, and the ratio of the smallest angular resolution to the Einstein angular radius (see, e.g., Dominik (2010) for description of the parameters). Some other input parameters, from which the above parameters can be derived, are also considered. The algorithm extracted the above parameters from input files obtained in the analysis of observations of microlensing events.

The developed algorithm finds the time intervals, in which different microlensing events can be observed, and the time intervals, in which it is better to observe particular events. It also plots diagrams of temporal changes in the seeing, air mass, sky brightness, and flux of events selected for observations with a particular telescope.

## MODELS OF THE NIGHT-SKY BRIGHTNESS

To create an algorithm that optimizes the selection of observed microlensing events in order to maximize the efficiency of finding exoplanets, a part of our research was aimed at constructing a model of the night-sky brightness and the seeing (Ipatov and Horne, 2014). As a basis for this, we used the infrared observations with the OGLE telescope and the FTS, FTN, and LT performed in 2011. The seeing value characterizes the image distortions caused by turbulence in the atmosphere. This night-sky model with different parameters for different telescopes is one of the elements of the algorithm to compare the capabilities of telescopes. It is based on the results of Krisciunas and Schaefer (1991).

To construct the night-sky model, the following observations were analyzed. For the OGLE telescope, we considered 20 events (110251–110270). For the FTS, FTN, and LT, we considered events observed in 2011, for which the size of data files is larger than 1 Kbyte (i.e., the minimal number of points in the light curve exceeds 15), namely, 20, 19, and 39 events for the LT, FTN, and FTS, respectively.

Unlike previous papers about the sky brightness (Benn and Ellison, 1998; Duriscoe et al., 2012; Krisciunas and Schaefer, 1991; Patat, 2003, 2008), we analyzed the observation data from different telescopes and used the $\chi^2$ optimization for this. The studies of the night-sky brightness may be of interest to different observers. For example, Ipatov and Elenin (2017a, 2017b) noted the importance of including the night-sky model into models aimed at determining the probability of detecting near-Earth objects in different regions of the sky.

### Dependence of the Seeing on the Air Mass

In Fig. 3 we show an example of the dependence of the seeing (expressed in arcseconds) on the air mass (which is $a \approx \sec(z)$, where $z$ is the zenith distance, i.e., the angle between the direction to the star and the zenith). The seeing is a quantity that characterizes the blurring of an image due to turbulence in the Earth's atmosphere and restricts the angular resolution of astronomical observations. It is measured as the width of the blurred disk at half maximum. In the distribution, the full width at half maximum (FWHM) is the difference between two values of the independent variable, at which the dependent variable is equal to half of its maximal value. Modern large optical telescopes include adaptive optics that prevents this blurring. A value less than $0.4''$ is considered excellent seeing. The air mass is a measure of the amount of air along the line of sight when observing a star or another celestial source from beneath the Earth's atmosphere. It is considered to be the integral of the air density along the line of sight. At zenith, by definition, it is equal to 1. In Fig. 3, the thick straight line is based on the $\chi^2$ optimi-zation ($s = 1.334 + 0.519(a - 1)$). This line characterizes the increase in the seeing $s$ with increasing air mass $a$ better than the considerations of the mean and median values of the seeing, which are represented by jogged lines in Fig. 3 and can be very different at close values of the air mass due to small statistics of the observations.

**Table 1.** The values of $s_0$, $s_1$, and $\sigma$ obtained with the $\chi^2$ optimization of the straight-line fit for the seeing $s$ in dependence on the air mass $a$ for observations with different telescopes

| Telescope | $s_0$ | $s_1$ | $\sigma$ |
|---|---|---|---|
| OGLE | 1.33 | 0.29 | 0.25 |
| LT | 1.35 | 0.42 | 0.50 |
| FTN | 0.68 | 0.21 | 0.21 |
| FTS | 1.33 | 0.52 | 0.37 |

Table 1 contains the values of $s_0$, $s_1$, and $\sigma$ obtained with the $\chi^2$ optimization of the straight-line fit for the seeing $s$ (the FWHM expressed in arcseconds) in dependence on the air mass $a$ ($a \approx \sec(z)$, where $z$ is the zenith distance): $s = s_0 + s_1(a - 1)$ ($\chi^2 = \Sigma[(s_j - s_1(a_j - 1) - s_0)/\sigma_s]^2$, where $\sigma_s^2$ is the dispersion and the sum is taken for the considered values of $s_j$ at air mass $a_j$). For the FTN, the seeing values $s$ were typically almost two times less than for the other three telescopes considered (OGLE, FTS, and LT). For these three telescopes, the typical seeing values $s$ were roughly the same, if the air mass is $a = 1$. However, at $a = 2$, they were smaller for the OGLE telescope than for the FTS and LT.

### Dependence of the Sky Brightness on the Air Mass

In Table 2, for the considered model of the night-sky brightness, we present the values of the zenithal brightness of the night sky $I_{sky}(0)$ for the case when the Moon is below the horizon, as well as the coefficients $b_o$, $b_{1o}$, and $b_1$, characterizing the dependence of the brightness $b$ on the air mass $a$ (where the brightness is $b = b_{1o}(a - 1) + b_o$ or $b = b_1(a - 1) + b_{oi}$). All of these values are given in stellar magnitudes per arcsecond squared.

The coefficients $b_o$ and $b_{1o}$ shown in Table 2 were based on the $\chi^2$ optimization of the straight-line fit ($b = b_{1o}(a - 1) + b_o$ and $\chi^2 = \Sigma[(b_j - b_{1o}(a_j - 1) - b_o)/\sigma_b]^2$, where $\sigma_b^2$ is the dispersion, $a$ is the air mass, $b_j$ is the brightness observed at the air mass $a_j$ when the Moon is below the horizon). The values of $I_{sky}(0)$, $b_o$, and $b_{1o}$ are presented for a value of 0.05 for the absorption coefficient (when this coefficient was 0 or 0.1, the $I_{sky}(0)$ values differed by less than 0.3%). The brightness $I_{sky}(0)$ was $18.1^m$ per arcsecond squared for the

**Table 2.** The night-sky brightness $I_{sky}(0)$ at zenith and the coefficients $b_o$, $b_{1o}$, and $b_1$

| Telescope | $I_{sky}(0)$ Moon is below the horizon | $b_o$ Moon is below the horizon | $b_{1o}$ Moon is below the horizon | $b_1$ Moon is below the horizon | $b_1$ all observations | $b_1$ the Sun elevation is $\theta_{Sun} < -18°$ | $b_1$ Moon is below the horizon and $\theta_{Sun} < -18°$ |
|---|---|---|---|---|---|---|---|
| OGLE | 18.1 | 18.0 | −0.22 | −0.24 | −0.24 | −0.24 | −0.23 |
| LT | 19.6 | 19.0 | −0.11 | −0.26 | −0.84 | −0.88 | −0.26 |
| FTN | 18.7 | 18.3 | −0.13 | −0.18 | −0.26 | −0.18 | −0.22 |
| FTS | 19.0 | 18.8 | −0.14 | −0.21 | −0.13 | −0.17 | −0.11 |

The values in the table are given in stellar magnitudes per arcsecond squared.

OGLE telescope, which exceeded the values for the three other telescopes considered. In this paper, all values of the sky brightness are given in stellar magnitudes per arcsecond squared, though the words "per arcsecond squared" are sometimes omitted.

Using the $\chi^2$ optimization of the straight-line fit $b = b_{1o}(a - 1) + b_o$ and considering a single value of $b_1$ for all microlensing events and different values of $b_{oi}$ for different events, we calculated the interval length $\Delta b = \max\{b_{oi}\} - \min\{b_{oi}\}$ for the $b_{oi}$ values. The $b_{oi}$ values characterize the zenithal brightness of the sky in the vicinity of different events. For observations carried out when the Moon was below the horizon, $\Delta b \approx 1^m$. The examples of the diagrams for the brightness of the sky in dependence on the air mass, including the lines $b = b_1(a - 1) + b_{oi}$, are shown in Figs. 4 and 5. As in the case of Fig. 3, with the $\chi^2$ optimization of the straight-line fit, the dependence of the night-sky brightness on the air mass is seen better. The dependences of the brightness and the seeing on the air mass, approximated by straight lines, are used in the algorithm to estimate the efficiency of detecting exoplanets with different telescopes. For the Moon below the horizon, the $b_{oi}$ values (characterizing the zenithal brightness of the sky in the vicinity of different events) usually differ, but no more than by $1^m$.

### Influence of the Positions of the Moon and the Sun on the Sky Brightness

For the different considered positions of the Moon and the Sun, the maximal values of $b_{0i}$ (the sky is less bright) are almost the same as the values for the Moon below the horizon. The difference from the min$\{b_{0i}\}$ values (the sky is brighter) can reach $1.5^m$ for the different positions of the Moon and the Sun. The $b_1$ values are presented in Table 2 for four constraints on the positions of the Moon and the Sun. Most of the observations were made for the air mass $a \leq 3$. The change in the sky brightness at $1 \leq a \leq 3$ does not exceed $|2b_1|$. For most values of $b_1$ presented in Table 2, $|2b_1| < 0.5^m$.

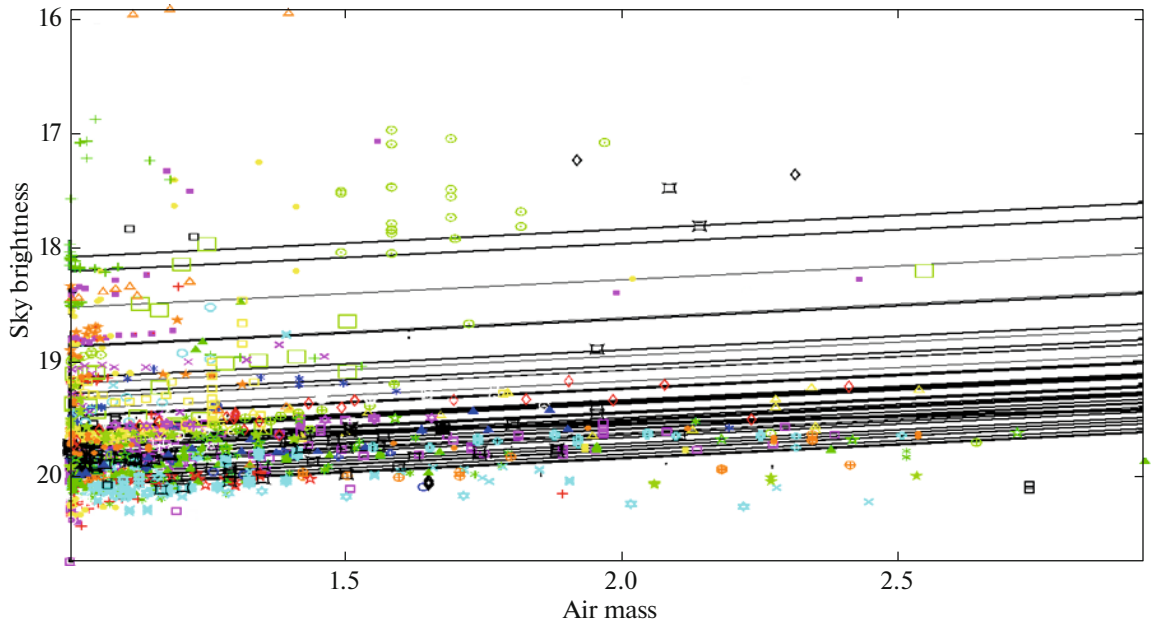

**Fig. 4.** The dependence of the night-sky brightness (expressed in stellar magnitudes per arcsecond squared) on the air mass for all positions of the Moon and the Sun in the FTS observations. The different lines are based on observations of 39 microlensing events. The lines were obtained for the $\chi^2$ optimization ($b = b_1 \times a + b_{oi}$) with the different values of $b_{oi}$ for different events and with the same value of $b_1$. The thick solid line was obtained for the model, in which $b_o$ takes the same value for all events.

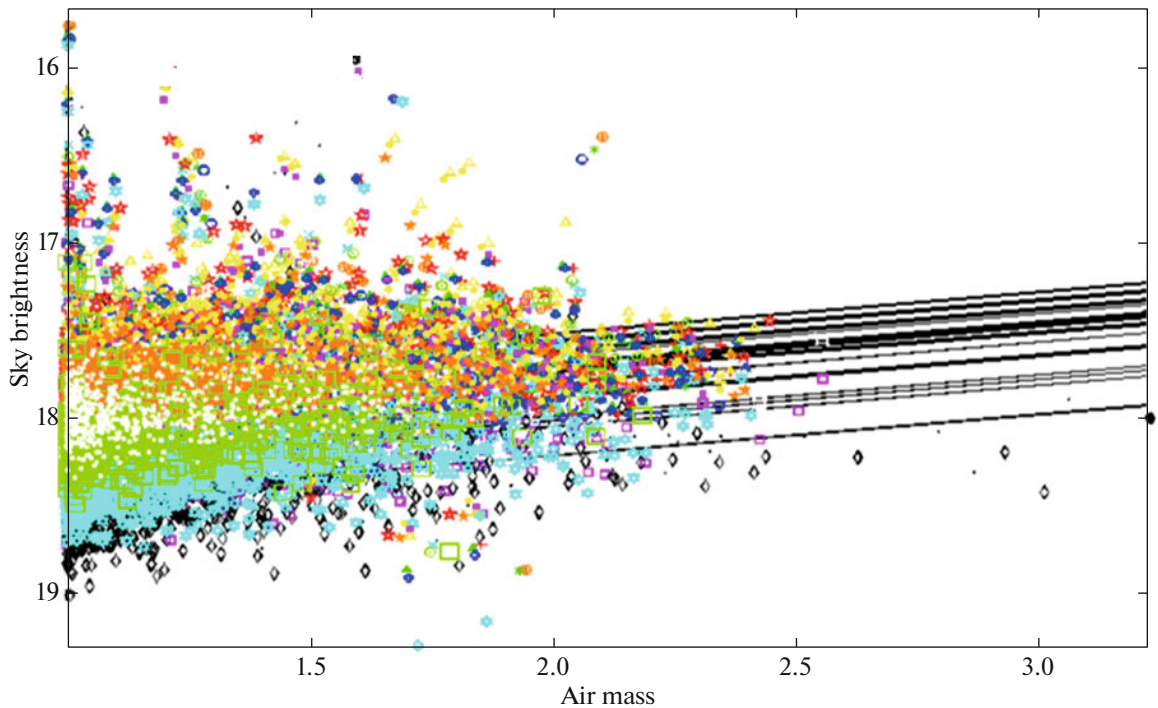

**Fig. 5.** The dependence of the night-sky brightness (expressed in stellar magnitudes per arcsecond squared) on the air mass for the OGLE observations, when the Moon was below the horizon of the Sun. The different lines are based on observations of 20 microlensing events. The lines were obtained for the $\chi^2$ optimization ($b = b_1 \times a + b_{0i}$) with the different values of $b_{0i}$ for different events and with the same value of $b_1$. The thick solid line was obtained for the model, in which $b_0$ takes the same value for all events.

For infrared observations with the FTS, FTN, LT, and OGLE telescopes, the zenithal sky brightness equaled to $19.0^m$, $18.7^m$, $19.6^m$, and $18.1^m$ per arcsecond squared, respectively. When we analyzed the images for the Moon below the horizon and the Sun elevation $\theta_{Sun} < -18°$, the sky brightness, which was usually observed near different microlensing events (i.e., regions of the sky) at the same air mass, generally varied by less than $1^m$ (depending on the events). The analysis of observations shows that the sky brightness may vary up to $5^m - 6^m$, when the Moon is bright and the Sun is close to the horizon. These bright regions of the sky are poorly simulated with the present sky model, but it is better not to observe them often.

For each of the events, the deviations in the night-sky brightness from the best model are the differences between the observations and the $\chi^2$ optimization that varies from one event to another. Even for all positions of the Moon and the Sun, these deviations were small, mostly ranging from $-0.4^m$ to $0.4^m$. For the Moon being below the horizon, the absolute magnitudes of many of the deviations did not exceed $0.2^m$. However, in this case, some deviations from the best model could reach $4^m$. Such large values of deviations were mainly caused by different elevations of the Sun in different observations. The maximal values of the sky brightness deviations for observations made at the Sun elevation $\theta_{Sun} < -18°$ and the Moon below the horizon are usually several times smaller than for all observations (for the FTS, compare Figs. 6 and 7). The influence of the Sun elevation $\theta_{Sun}$ on the night-sky brightness became observable at $\theta_{Sun} > -14°$ and is significant at $\theta_{Sun} > -7°$. For example, in the FST observations, when the Moon was below the horizon, the deviations in the night-sky brightness $s_{br}$ exceeded $-0.4^m$ and $-1^m$ at $\theta_{Sun} < -14°$ and $\theta_{Sun} < -8°$, respectively; and the $s_{br}$ values may reach $-3^m$ at $-8° < \theta_{Sun} < -7°$ (see Fig. 8). For the considered observations, the sky was not bright (i.e., the lower limit in deviations was larger than $-1^m$) only in the case, when the Moon was below the horizon and the $\theta_{Sun}$ values were simultaneously less than $-18°$.

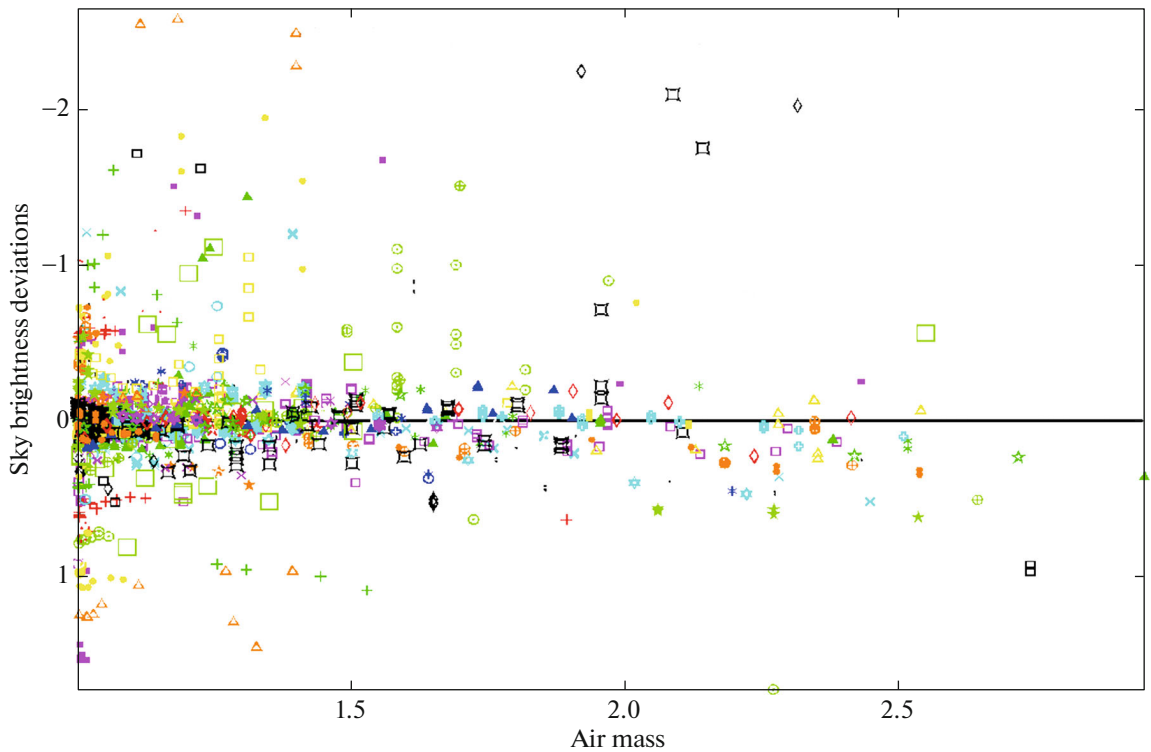

**Fig. 6.** Deviations in the night-sky brightness (expressed in stellar magnitudes) from the model in dependence on the air mass. The simulations were performed with different values of $b_{oi}$ for the FTS observations of 39 microlensing events. All positions of the Moon and the Sun were considered.

## SELECTION OF MICROLENSING EVENTS FOR OBSERVATIONS

The algorithm we developed allows us to increase the efficiency of detecting exoplanets in observations. It can be used to identify, for a given telescope, those of the known microlensing events that are better to observe at the time of interest to improve the efficiency. Figure 9 shows an example of the time intervals for the events selected for the FTS observations.

To estimate the comparative efficiency of detecting exoplanets, for the "best" events (i.e., the events selected for observations at the current time point), we considered the value of $w_{sum} = \sum g_i[(\Delta t + t_{done})^{1/2} - t_{done}^{1/2}]$ (where $\Delta t = 2t_{slew}$ for the event observed at the current time moment, while $\Delta t = t_{slew}$ and $t_{done} = 0$ for the other events; the coefficients $g_i$ have been discussed above when describing the algorithm and the detection zone of the $i$th event) and $r_{wsumt} = (w_{sum}/w_{sumOGLE})/(t_{sum}/t_{sumOGLE})$, where $t_{sum}$ is the total duration of the specified time interval, for which at least one event can be observed ($t_{sumOGLE}$ is the value for the OGLE). The $r_{wsumt}$ values were computed and compared for 13 telescopes. For the best events, we also calculated $w_{sumo} = \sum g_i[(t_s + t_{done})^{1/2} - t_{done}^{1/2}]$ and $r_{wsumto} = (w_{sumo}/w_{sumoOGLE})/(t_{sum}/t_{sumOGLE})$ for $t_s = 20$ s. In the formula for $w_{sumo}$, it is always assumed that $t_s = 20$ s, while in the formula for $w_{sum}$ the values of $\Delta t$ may differ for different events. Usually, the $t_{sum}/t_{sumOGLE}$ value obtained in calculations is about 0.66, 0.56, and 0.62 for the FTN, LT, and MDO telescopes, respectively; however, it may vary for different time intervals considered. For the other telescopes of interest, this ratio differed from unity by less than 0.1.

For observations at the 1-m telescope (located at the CTIO, SAAO, SSO, or MDO) equipped with a Sinistro CCD camera and at the 2-m telescope (the FTS, FTN, or LT), our calculations of $r_{wsumt}$ and $r_{wsumto}$ (characterizing the capability of the telescopes with respect to the OGLE) mostly yielded the values in intervals of 0.8–1.2 and 1.4–2.2, respectively (see Fig. 10). The data shown in Fig. 10 are based on the analysis of 1562 microlensing events accessible for observations in 2011. The data for the intervals that started on April 22 and August 1 are marked with crosses and ellipses, respectively (Fig. 10). The data marked with large black symbols were obtained for real values of the time moments $t_0$ corresponding to the peaks of the brightness curves. The data obtained at random values of $t_0$ are marked with small red symbols. The data for real values of $t_0$ and the 90-day interval beginning on April 22 are marked with green crosses. The remaining data are for the 100-day interval. In the calculations, the results of which are presented in

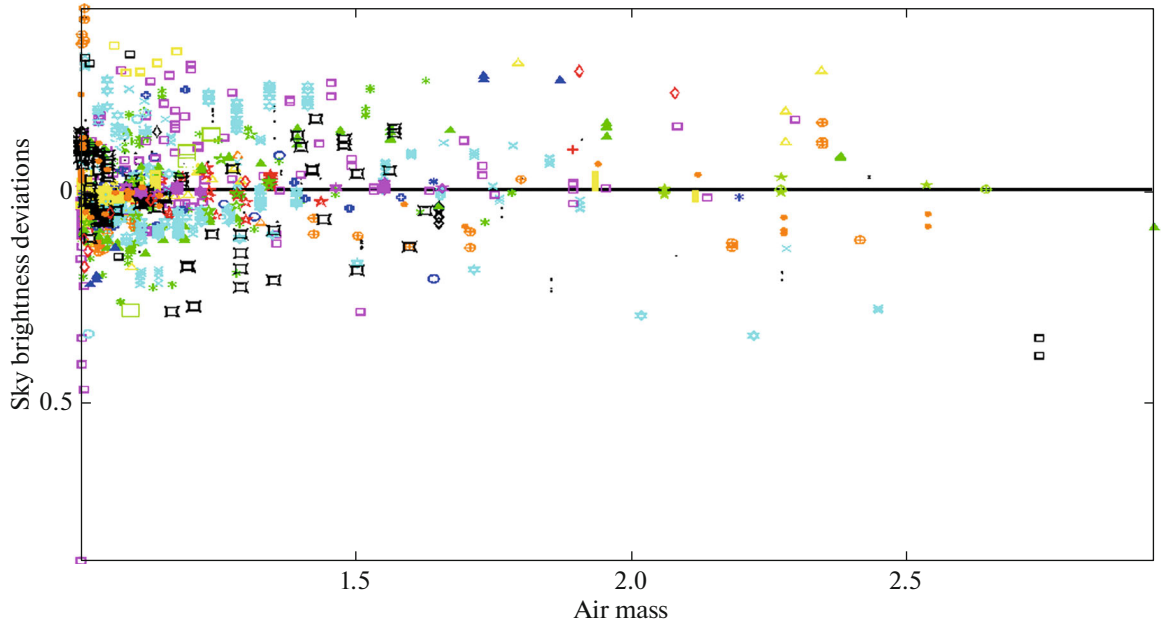

**Fig. 7.** Deviations in the night-sky brightness (expressed in stellar magnitudes) from the model in dependence on the air mass. The simulations were performed with different values of $b_{oi}$ for the FTS observations of 39 microlensing events. The cases for the Moon below the horizon and the elevation of the Sun at $\theta_{Sun} > -18°$ were considered.

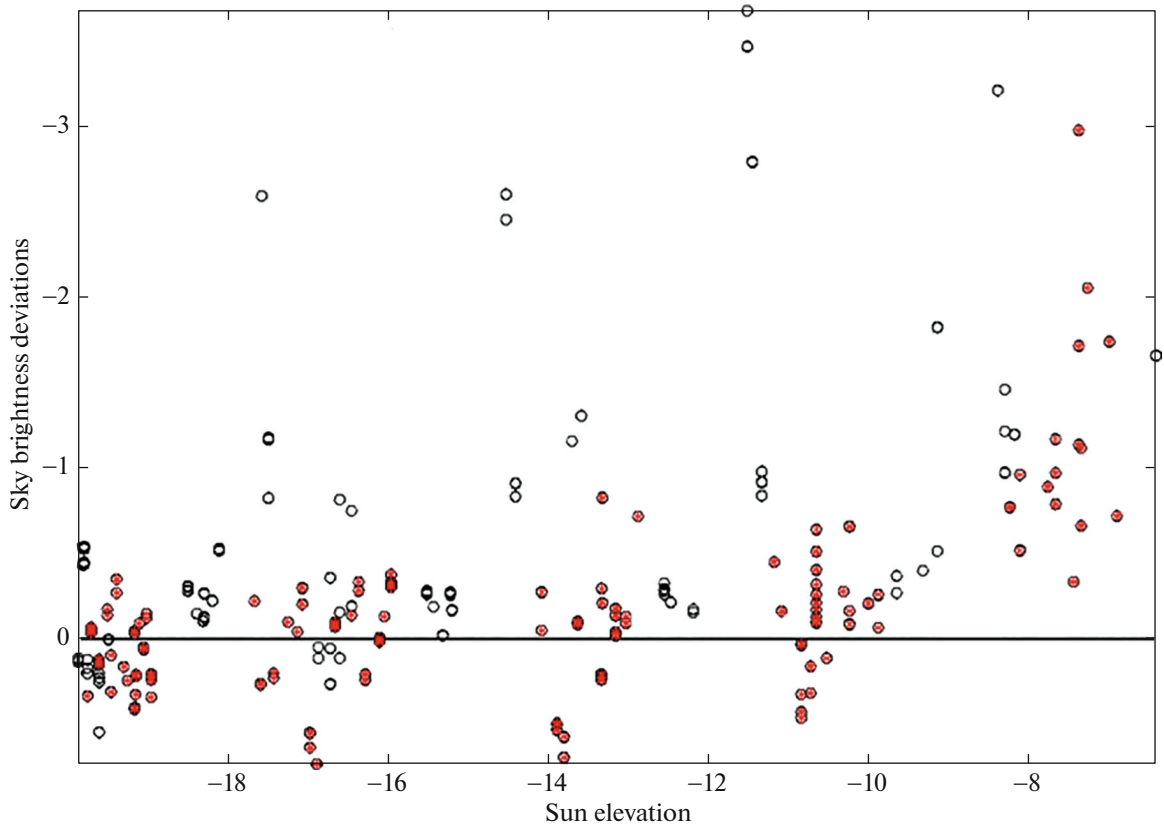

**Fig. 8.** Deviations in the night-sky brightness (expressed in stellar magnitudes) from the model in dependence on the elevation of the Sun $\theta_{Sun}$ (expressed in degrees). The simulations were performed with different values of $b_{oi}$ for the FTS observations of 39 microlensing events. The circles correspond to all of the observations. The red stars indicate that the Moon was below the horizon.

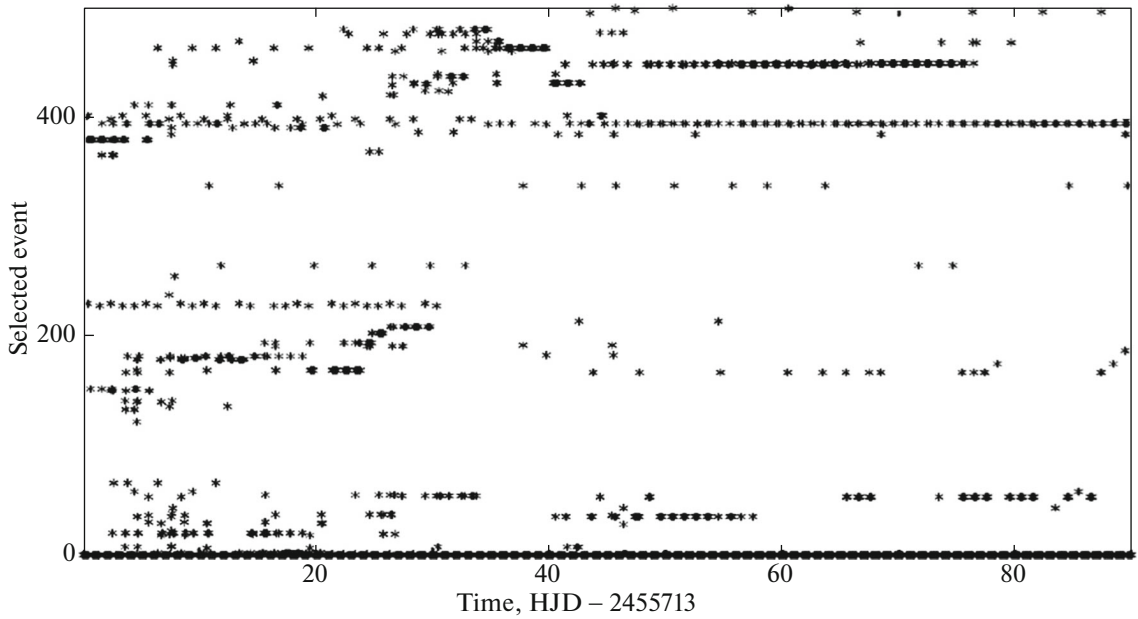

**Fig. 9.** Time intervals (expressed in days) for the events selected to be observed with the FTS telescope (at real moments of peaks in the light curves). The analyzed microlensing events (110 501−111 000) are numbered from 1 to 500 along the ordinate axis. The zero event corresponds to the case when there were no observations.

Figs. 10 and 11, it was assumed that the 1-m telescopes (with the Sinistro CCD camera), located at the same site, observed different events at the same time. The non-priority telescopes, the data for which are marked with small symbols in Fig. 10, were not allowed to observe the events that were already observed at that moment with the priority telescopes at the same observatory. The calculations were performed both for real values of the time moment $t_0$ corresponding to the peak of the brightness curve and for random $t_0$ values of the beginning of the interval ($t_0 = R_{NDM}(t_{mx} + 2t_E) - t_E$, where $R_{NDM}$ is a random value from 0 to 1, $t_{mx}$ is the duration of the interval considered, and $t_E$ is the time equal to the ratio of the Einstein angular radius to the corresponding relative velocity). When the $t_0$ moments were randomly selected, the number of peaks in the brightness curve was larger than for the real values of $t_0$.

For the FTS and the SSO telescopes, which are located in the same place but differ by a factor of two in diameter of the mirrors, the ratio of the $w_{sum}$ values was usually about 2. The $w_{sum}$ (and $w_{sumo}$) values were smaller by a factor of about 1.2 for the SBIG CCD camera than for the Sinistro CCD one. For observations at the 1-m telescope with the Sinistro CCD camera, this efficiency was often about 0.8 of the value for the OGLE telescope. However, in some cases, it was larger than the efficiency for OGLE. The difference in the $w_{sum}$ values is about 5%, if we used the same $I_{sky}(0)$ values and the same dependence of the seeing on the air mass for the telescopes at the SSO and the OGLE telescope compared to using the same values and dependence for the FTS. In contrast to Fig. 10, which shows only the general data for all of the observations analyzed, Fig. 11 allows us to compare the telescopes in terms of the efficiency of detecting planets by microlensing for different groups of events and different observation intervals. In Fig. 11, in the considered time interval $T = 90$ d (i.e., within this interval, the number of peaks in the light curve is large), the black and green symbols correspond to the real values of $t_0$ (at the peaks in the light curve) and the random values of $t_0$ ($t_0 = R_{NDM} \times 2t_E - t_E$, where $t_E$ is the time scale equal to the ratio of the Einstein angular radius to the relative proper motion), respectively; the red and blue symbols correspond to the time interval $T = 2$ d (the number of peaks is small).

In the model we have considered, the search for exoplanets is based on already known microlensing events. The capabilities of telescopes in search for exoplanets were compared by simulations under the assumption that each telescope observes the microlensing event, for which the calculated exoplanet detection probability is maximal among the events analyzed. The results of comparison show that, for the 2-m FTS, FTN, or LT, the comparative $r_{wsumt}$ efficiency of detecting exoplanets per unit time is higher (typically, by a factor of 1.4−2.2) than for the OGLE telescope. The probability of detecting an exoplanet is usually proportional to the mirror diameter of a telescope.

The ratio $r_{50}$ of the $w_{sum}$ value (characterizing the efficiency of detecting an exoplanet), corresponding to observations of only events with the maximal mag-nification in brightness $A_{max} > 50$ (4% of all events), to the $w_{sum}$ value, corresponding to the case when all 1562 events can be observed (the maximal magnification in the brightness of a star $A_{max}$ may take any value), can differ by an order of magnitude for different time intervals specified. For example, for the OGLE observations and the 5-day intervals beginning on April 22 and August 1, 2011, $r_{50}$ is 0.07 and 0.99, respectively. For the 100-day intervals, $r_{50}$ is 0.63 and 0.83, respectively. For $50 < A_{max} < 200$ (2% of all events), the analogous values of $r_{50-200}$ were 0.056, 0.78, 0.09, and 0.50 respectively. For the considered telescopes and the 100-day time intervals, the values of $r_{50}$ and $r_{50-200}$ were within 0.58–0.74 and 0.06–0.21, respectively, if the time intervals began on April 22, and within 0.73–0.88 and 0.29–0.52, respectively, if the time intervals began on August 1, 2011. The $w_{sum}$ values derived with our algorithm are close to the values obtained in the selection of events for observations by Dominik et al. (2010).

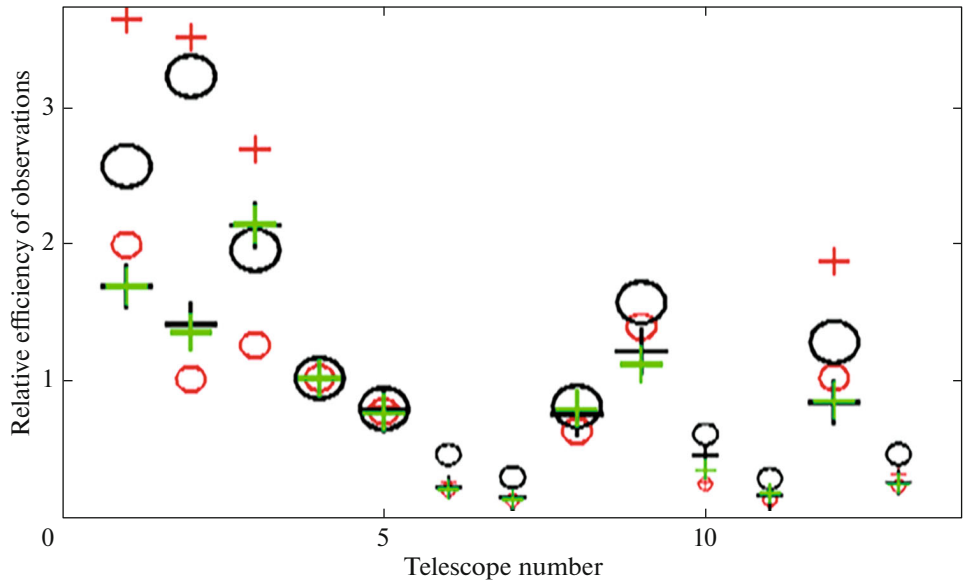

**Fig. 10.** The ratio of the exoplanet detection probability for the considered telescope to the analogous probability for the OGLE telescope $r_{wsumt} = (w_{sum}/w_{sumOGLE})/(t_{sum}/t_{sumOGLE})$ in dependence on the telescope number. The estimates are based on the analysis of 1562 microlensing events occurred in 2011. See the text for designations.

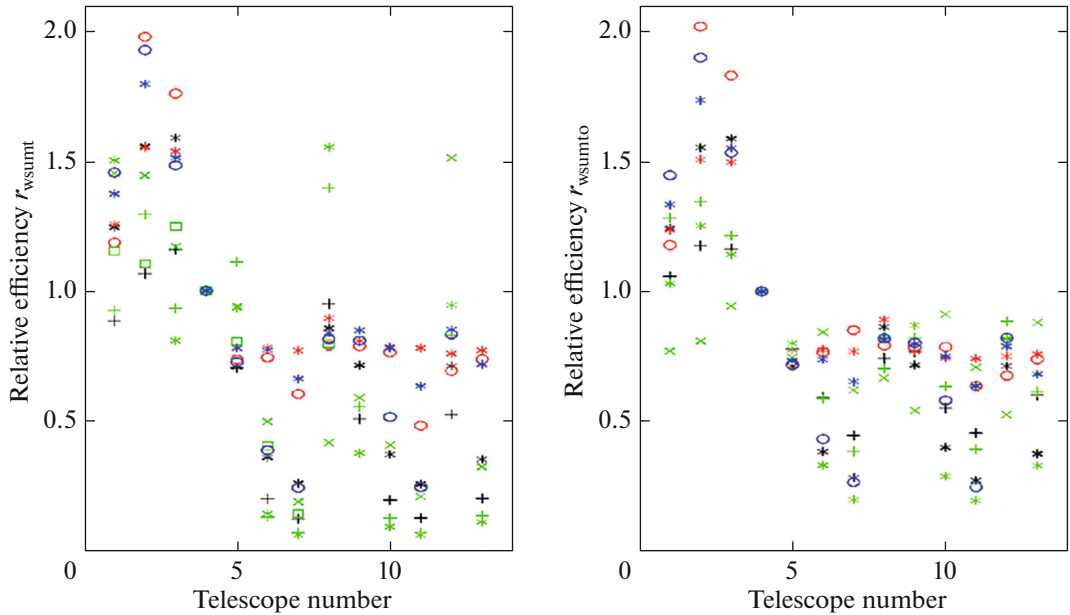

**Fig. 11.** The values of $r_{wsumt} = (w_{sum}/w_{sumOGLE})/(t_{sum}/t_{sumOGLE})$ (left) and $r_{wsumto} = (w_{sumo}/w_{sumoOGLE})/(t_{sum}/t_{sumOGLE})$ (right) characterizing the efficiency of observations of microlensing events in dependence on the telescope number. The data for events with numbers in ranges of 110001–111562, 110501–111000, 111001–111500, 110251–110400, and 110261–110270 are marked by symbols □, +, ×, *, and O, respectively. The colors of symbols are explained in the text.

The OGLE telescope observed a little more than 200 galactic fields. For 1500 events observed in 2011, this means that there could be about 10 events (the 1500/200 ratio is 7.5) in one field. For the other telescopes (except for the LT), only one event can be found in the field of view. In our calculations for 1562 events (and a time interval of ~100 days), the $w_{sum}$ (or $w_{sumo}$) value for the best events selected for observations was usually much larger than for the other ten typical events (which are not the best at the current moment). Hence, we can use the $w_{sum}$ (or $w_{sumo}$) values calculated only for the best events to compare the capability of different telescopes in search for new exoplanets.

Our estimates are based on the data from 1562 microlensing events (with numbers 110001–111562) already detected by the OGLE in 2011. For a 90-day (or longer) period of observations, the main (usually more than half) contribution to the exoplanet detection probability (in terms of $w_{sum}$) came from the observations performed during short time intervals that included peaks in the light curve, if the telescope was allowed to observe all of the events. Depending on the presence of peaks in the light curve during the observations, the efficiency for the $r_{wsumt}$ telescope relative to the efficiency for the OGLE telescope can differ by a factor of up to 3.

To search for new microlensing events, it is better to use telescopes with a wide field of view, such as the OGLE telescope. However, our evaluations suggest that, for observations of known events, the efficiency of searching for exoplanets with the 1-m telescope with the Sinistro CCD camera can be close to that for the OGLE telescope, while the 2-m RoboNet telescope is on average more efficient than the OGLE one. The ratio of the $w_{sum}$ values (or the $w_{sumo}$ values) per unit of time for the 2-m telescope to the value for the OGLE telescope was usually in a range of 1.4–2.1, while for the 1-m telescope this ratio was usually in a range of 0.8–1.2. It is believed that, in searching for exoplanets with the microlensing method, the greater efficiency for telescopes with a larger diameter of the mirror may be, in particular, due to the fact that a telescope with a larger diameter requires less exposure time, and more images can be taken overnight, while the peak in microlensing on an Earth-type planet may last only 1–2 h.

In our test calculations, the $w_{sum}$ values (or the $w_{sumo}$ values) were usually two times larger for a telescope with the 2-m diameter than for a 1-m telescope, if the telescopes differed only by the diameter $d$ of the mirror. This factor corresponds to $s_{ef}^{0.5}$, where $s_{ef}$ is the effective area of a telescope. For two or three telescopes observing the same event from the same site, the ratio of $w_{sum}$ (or $w_{sumo}$) to the corresponding value for a single telescope is about $2^{0.5} \approx 1.4$ or $3^{0.5} \approx 1.7$, respectively.

Based on the above results, we may conclude that it is better to use nearby telescopes for observations of different microlensing events. However, it is often better to use all such telescopes to observe the same event at the time points corresponding to the brightness peak of the event. The algorithm and night-sky brightness models we developed can also be used to plan a variety of observations, including not only microlensing events.

## CONCLUSIONS

We compared the efficiency of detecting exoplanets (the probability of detection per unit time) in observations of microlensing events with 13 different telescopes and with several approaches to the selection of observable events. In constructing the algorithm to select optimal targets for these observations and in comparing the capabilities of several telescopes, we considered models of the night-sky brightness that satisfy infrared observations with the OGLE telescope and the RoboNet telescopes (FTS, FTN, and LT) carried out in 2011 to search for planets by microlensing. The time intervals, during which the events can be observed, were determined with accounting for the positions of the Sun, the Moon, and the other constraints on the telescope pointing. In particular, the dependences of the night-sky brightness and the seeing on the air mass were considered.

Our algorithm allows us to identify the known microlensing events available for observations with a particular telescope and to select targets, for which the probability of detecting exoplanets is maximal. For these estimates, we used data on 1562 microlensing events already detected with the OGLE telescope in 2011. The potential of telescopes to search for exoplanets was compared by simulations under the assumption that each telescope observes the microlensing event, for which the calculated exoplanet detection probability is maximal among the events considered. To search for new events, it is better to use telescopes with a wide field of view, such as the OGLE one. However, our estimates showed that, for observations of known events, the efficiency of searching for exoplanets with the 1-m telescope equipped with the Sinistro CCD camera can be close to the efficiency for the OGLE telescope, and the 2-m RoboNet telescope is typically 1.4–2.1 times more efficient than the OGLE one. The relative probability of detecting exoplanets with a particular telescope compared to the OGLE one depends on the presence of peaks in the light curve during observations. The probability of detecting exoplanets is usually proportional to the mirror diameter of a telescope.

Based on our results, we can conclude that nearby telescopes are usually better used to observe different microlensing events. However, it is often better to use all of these telescopes to observe the same event at the time moments corresponding to the brightness peak of the event. For a 90-day (or longer) period of observations, the major (usually more than half) contribution to the probability of detecting an exoplanet came from observations made during short time intervals that included peaks in the light curves, if the telescope was allowed to observe all of the events. The algorithm and the night-sky brightness models we developed can also be used to plan a variety of observations, including not only microlensing events.

## ACKNOWLEDGMENTS

The author is grateful to Keith Horne for supervising this research. He also would like to thank all of the co-authors of previous joint publications on this subject for collaborating on the problems of searching for exoplanets by the microlensing method. The author acknowledges the reviewer for useful remarks that contributed to improving the paper.

## FUNDING

This study was supported by ongoing institutional funding of the Vernadsky Institute of Geochemistry and Analytical Chemistry of the Russian Academy of Science.

## CONFLICT OF INTEREST

The author of this work declares that he has no conflicts of interest.